\documentclass{elsart}
\usepackage{epsfig}

\begin{document}

\begin{frontmatter}
\title{Complexity vs Stability in Small-World Networks}

\author{Sitabhra Sinha}

\address{The Institute of Mathematical Sciences, C. I. T. Campus,
Taramani, Chennai - 600 113, India.}

\begin{abstract}
According to the May-Wigner stability theorem,
increasing the complexity of a network inevitably leads to its
destabilization, such that a small perturbation will be able to disrupt
the entire system.
One of the principal arguments against this observation
is that it is valid only for random networks, and therefore does not
apply to real-world networks, which presumably are structured.
Here we examine how the introduction of small-world 
topological structure into networks affect their stability. Our results
indicate that, in structured networks, the parameter values at which the 
stability-instability transition 
occurs with increasing complexity is identical
to that predicted by the May-Wigner criteria. However,
the nature of the transition, as measured by the
finite-size scaling exponent, appears to change as the network topology
transforms from regular to random, with the small-world regime as the 
cross-over 
region. This behavior is related to the localization of the largest
eigenvalues along the real axis in the eigenvalue plain with increasing
regularity in the network.

\vspace{0.25cm}
\end{abstract}

\end{frontmatter}

\section{Introduction}
The issue of whether increasing the complexity of a network contributes
to its dynamical instability has long been debated. This `complexity vs
stability' debate is especially acute in the field of ecology \cite{McC00}, 
as it
relates to the importance of diversity for the long-term survival of 
ecosystems.
However, understanding the relation between the network structure and 
its stability
(with respect to dynamical perturbations) is crucial,
as it is related to the robustness of systems as ubiquitous as
power grids, financial markets, and even complex societies and 
civilizations\cite{Tai90}.
Pioneering studies on the stability of networks, both theoretical\cite{May73}
and
numerical\cite{Gar70}, suggested that increasing the network complexity,
as measured by its size ($N$), density of connections ($C$)
and the strength of
interactions between coupled elements ($\sigma$), almost inevitably leads to 
the destabilization of any arbitrary equilibrium state of the system.
This result, known as the May-Wigner stability theorem, seemed to fly in
the face of conventional wisdom that higher diversity makes a system
more capable of surviving perturbations and has since led to much research
on the connection between network complexity and stability\cite{Wil02}.

\vspace{-0.3cm}
The May-Wigner argument \cite{May73} confines itself to analyzing the local
stability of an arbitrarily chosen equilibrium point of the network dynamics.
Under such constraints, the explicit dynamics at the nodes can be ignored
and the stability is governed by the leading eigenvalue of the linear
stability matrix ${\bf J}$. As a first approximation, one can consider
the network elements to be coupled randomly with each other. If the connection 
weights between linked nodes follow a Gaussian distribution (with mean 0 and
variance $\sigma^2$), then it follows that ${\bf J}$ is a random matrix.
Therefore, existing rigorous results on the eigenvalue distribution of
random matrices can be applied, which allows one to make the assertion
that the network is almost certainly stable if $\sqrt{N C \sigma^2} < 1$, 
and almost certainly unstable otherwise.

\vspace{-0.3cm}
Objections to the May-Wigner argument have often revolved around the assumption 
of a randomly connected network. As pointed out by many ecologists, most
networks occurring in nature are not random, and seem to have structures
such as trophic levels in the predator-prey relations between
different species.
Some early studies seemed to suggest that introducing a hierarchical
organization (e.g., by partitioning the adjacency matrix of the network
into blocks \cite{McM75} or by having tree structures \cite{Hog89})
can increase the stability of a network under certain conditions. 
However, no general consensus on this issue has yet been achieved. 

\vspace{-0.3cm}
The introduction of ``small-world" connection topology\cite{Wat98}
has allowed the possibility of having different kinds of structures
in a network, other than a straightforward hierarchy of levels.
Small-world networks have the global properties of a random network
(short average path length between the elements) while at the local
level they resemble regular networks with a high degree of clustering
among neighbors.
In fact, several empirically obtained food web networks have been analyzed
by different research groups looking for evidence of small-world structure.
Initial reports of small-world ecological networks based on the analysis of
4 food webs \cite{Mon02} have been challenged by a study based on
7 food webs \cite{Cam02}, and, more recently, by a comprehensive analysis
of 16 food webs covering a wide variety of habitats\cite{Dun02}. 
The latter studies did not see significantly high clustering in most 
of these systems, compared to a random network.

\vspace{-0.3cm}
In light of this, it is inevitable to ask oneself whether the introduction
of small world connectivity confer any advantage to the network. If the
occurrence of higher than average clustering has no functional significance,
then the occurrence of small-world structures in a few networks are
probably due to chance alone. In particular, we can ask whether introducing
such structures in a network can make it more stable, and therefore,
able to survive perturbations compared to its random counterpart.

\vspace{-0.3cm}
In this paper we strive to answer the above question. Although there
have been previous studies on the eigenvalue distribution of small-world
networks\cite{Mon99,Far01}, the issue of stability has not been looked at
in any depth. In the next section,
we have described the basic model used to study the stability of the network
as its structure is changed from regular through small-world to random. The
results of extensive numerical studies is reported in Section 3, which suggests
that the stability-instability transition occurs at the same critical value 
independent of the network structure; but the nature of this transition
(as measured by the finite-size scaling exponent)
changes with the topology. 
Finally, we conclude with a brief discussion of the
implications of our results.

\section{The Randomly Weighted Network Model}
\vspace{-0.3cm}
To observe the stability of networks at the small-world regime
we follow the basic Watts-Strogatz construction \cite{Wat98}. A ring
consisting of $N$ nodes, with each node connected to $2 k $ neighbors
(i.e., neighborhood size is $k$), is rewired with probability $p$.
In other words, a fraction $p$ of the links among the nodes in the lattice are 
broken and then randomly reconnected, subject to the condition that 
the total number of links
does not change and that two nodes are not connected by more than one
directed link.
As outlined in Ref.\cite{Wat98}, increasing $p$ decreases both the
average path length and the clustering between nodes.
In addition, we introduce randomly distributed
weights to each of the links. Following May \cite{May73}, we generate
the corresponding linear stability matrix ${\bf J}$, such that, the
non-zero entries are chosen from 
a Gaussian distribution with mean 0 and variance $\sigma^2$. 
In addition, to ensure that the nodes are individually stable in the
absence of connections, the diagonal elements of ${\bf J}$
are chosen to be $-1$.
For a given
$p$, the stability
of the resulting network is then examined by observing the sign of the
leading eigenvalue $\lambda_{max}$ of ${\bf J}$ as a function of size ($N$),
connectance ($C \simeq {\frac{2k+1}{N}}$) and 
strength of connectivity ($\sigma$).
If $\lambda_{max} > 0$, the network is considered unstable. The
corresponding probability of stability $P_{stability}$ 
is calculated by carrying
out a large number of network realizations. While this analysis does not
explicitly consider dynamics, recent studies\cite{Che01,Sin04} indicate
that including the dynamics of the nodes does not qualitatively change 
the results obtained using the above technique. 

\vspace{-0.3cm}
It is possible that a sparsely connected network can be broken up into
disconnected clusters by the rewiring procedure. 
For this reason, in the simulations reported
here we have used $k \gg {\rm ln} (N)$, which ensures that the
entire network remains connected.
Note that, unlike another study on the stability 
of small-world
networks \cite{Li03}, we are analyzing the stability of
an asymmetric sparse matrix whose non-zero entries are normally distributed.

\section{Results}
\vspace{-0.3cm}
As mentioned in the previous section, we observe the order parameter
$P_{stability}$ against the different network complexity parameters.
For instance, keeping $N$ and $C$ fixed, as we increase $\sigma$,
$P_{stability}$ decreases from 1 to 0, i.e., the
network shows a stability-instability transition.
The critical parameter value, $\sigma_c$, at which the transition occurs 
remains
unchanged as we vary the connection topology from regular ($p = 0$)
through small-world to 
random ($p = 1$). This implies that changing the connection topology 
does not affect the stability of a network. However, the transition appears
to get sharper 
as the network becomes more random.

\begin{figure}
\epsfxsize=5.5in
\centerline{\epsffile{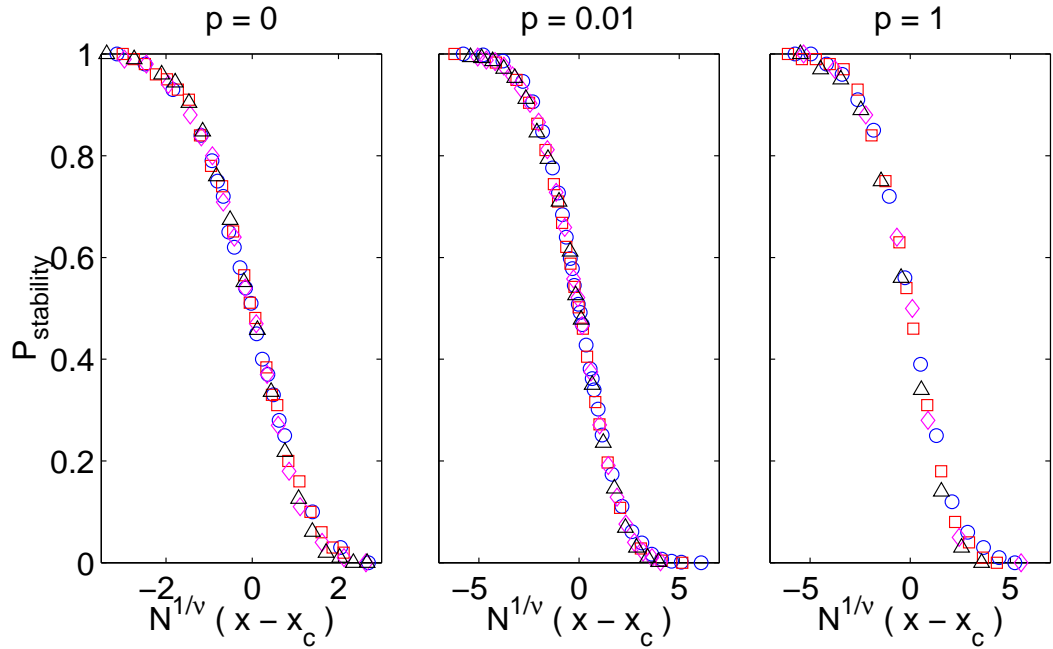}}
\caption{Finite-size scaling of the stability-instability transition
for different network topologies ($C \simeq 0.1$). 
The tuning parameter is $x 
= \sqrt{N C \sigma^2}$ ($x_c \rightarrow 1$ as $N \rightarrow \infty$, as
predicted by the May-Wigner theorem) 
and the order parameter is $P_{stability}$, the 
probability that a network is stable (i.e., $\lambda_{max} < 0$).
The scaling exponent $\nu \simeq 2$ for
$p = 0$ (left), $\nu \simeq 1.72$ for $p = 0.01$ (center) and $\nu \simeq
1.5$ for $p = 1$ (right). Data shown for $N = 200$ (circles), $N = 400$
(squares), $N = 800$ (diamonds) and $N = 1000$ (triangles). 1000 realizations
were performed for each data point.
}
\label{fig0}
\end{figure}

\vspace{-0.3cm}
To quantitatively measure the increase in steepness with randomness we 
used finite-size scaling analysis. The sharpness of the stability-instability
transition increases with $N$ for all topologies; 
finite-size scaling allows us to measure
how the relative width of the transition region decreases with increasing $N$.
It was noted by May\cite{May73} that for random networks, this scales
as $N^{-2/3}$. We have carried out this analysis for networks with 
different values of $p$, where the width of
the transition region scales as $N^{-1/\nu}$, and we observe
the variation of $\nu$ with $p$. As shown in Fig. 1, $\nu \simeq 2$ for
regular networks, and gradually decreases with $p$, ultimately 
becoming $\frac{3}{2}$ for $p = 1$ (as expected at the random network limit).

\begin{figure}
\epsfxsize=5.5in
\centerline{\epsffile{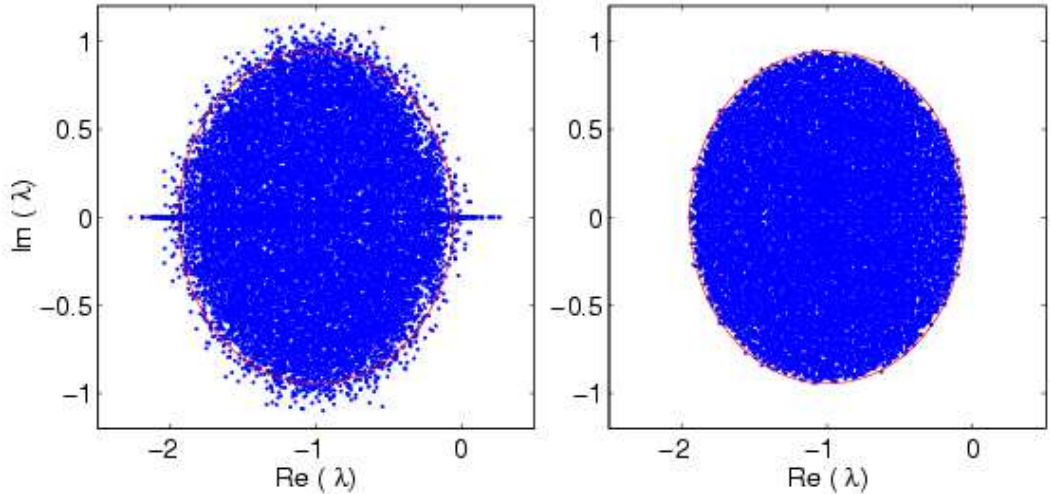}}
\caption{The eigenvalue plain for
regular (left) and random networks (right) with $N = 1000$, $C = 0.021$
and $\sigma = 0.206$. The data shown is for 20 realizations
of each kind of network. Note the tails along the real axis for $p = 0$.}
\label{fig2}
\end{figure}

\begin{figure}
\epsfxsize=5.5in
\centerline{\epsffile{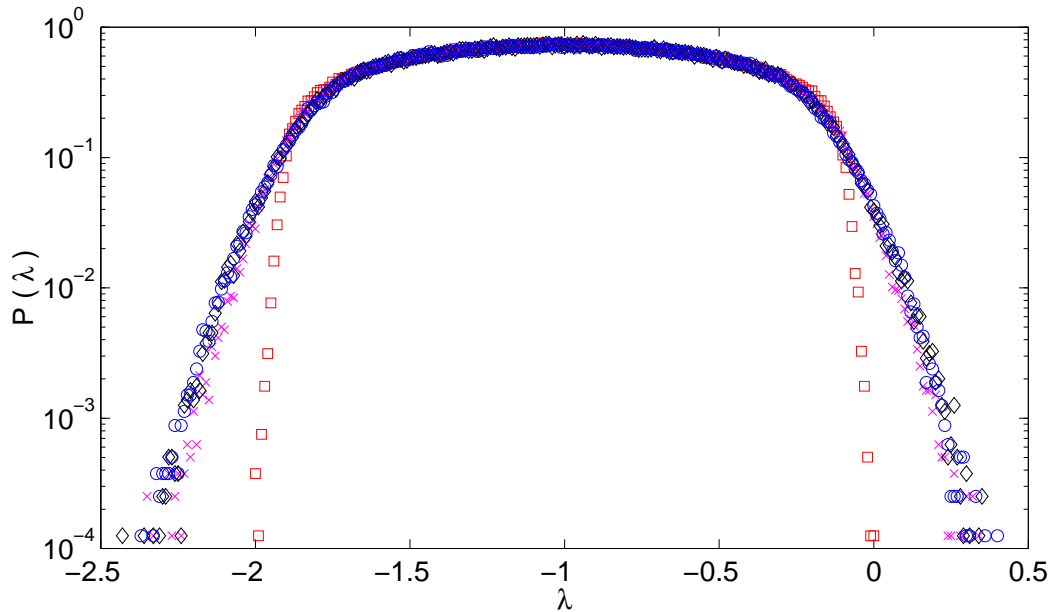}}
\caption{The eigenvalue distributions for networks with $p = 1$ 
(squares), $p = 0.1$ (crosses), $p = 0.01$ (diamonds) and
$p = 0$ (circles).
($N = 800$, $C = 0.021$ and $\sigma = 0.206$). The data points are
obtained after averaging over 1000 realizations.
}
\label{fig3}
\end{figure}

\vspace{-0.3cm}
To understand why the nature of the stability-instability transition is 
affected by the network topology, we look at the eigenvalue plain 
of the regular and random networks (Fig. 2). The eigenvalues of the
latter are bounded by a circle centered at $-1$ and having a radius of
$\sqrt{N C \sigma^2}$. However, for regular networks, there are
extensions from this circle along the real axis. The largest eigenvalues
are located on this `tail'. The extended tails of the eigenvalue 
distribution for $p < < 1$ are shown in greater detail in Fig. 3.
For $p = 1$, the distribution is bounded, as predicted by Wigner's
semicircle theorem. However, in the presence of clustering (i.e., as
$p \rightarrow 0$), the distribution extends out of the limits predicted
by the semicircle distribution. 
The stability of the network is governed by the maximal eigenvalue.
For the regular network, this is found at the
tail of the eigenvalue distribution where the relative variance 
of $\lambda_{max}$ is much larger than if it was located in the bulk 
(as is the case for
$p = 1$). This results
in a smoother transition from stability to instability for regular networks.

\begin{figure}
\epsfxsize=5.5in
\centerline{\epsffile{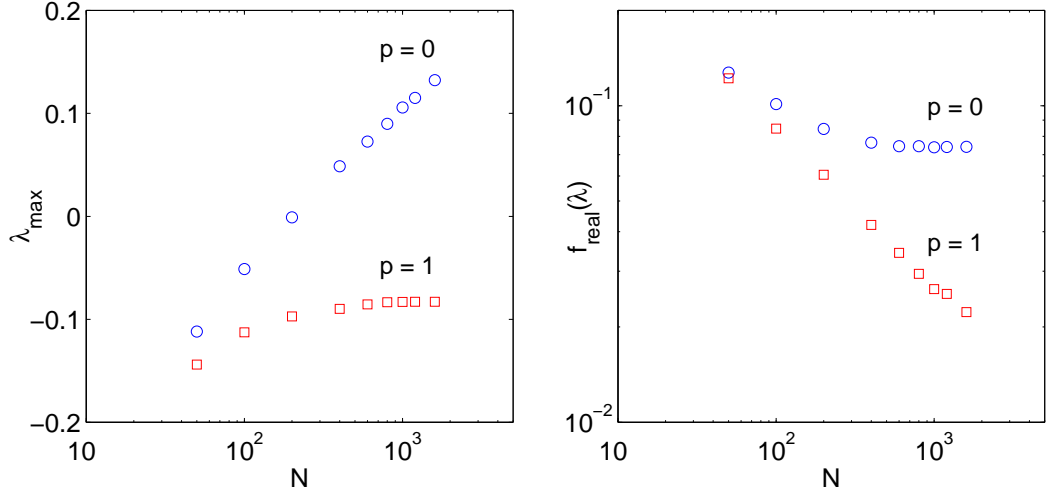}}
\caption{The largest eigenvalue $\lambda_{max}$ (left) and the fraction
of real eigenvalues $f_{real} (\lambda)$ (right) plotted against
network size $N$ for 
regular ($p = 0$) and random networks ($p = 1$) with
$k = 10$ and $\sigma = 0.206$. The data points are obtained by
averaging over 1000 realizations.}
\label{fig1}
\end{figure}
\vspace{-0.3cm}
It has been pointed out in Ref.~\cite{Far01} that in real-world networks,
links are `expensive'. Therefore, we also looked at the
case where $k$ is fixed as the system size increases (so that $C$ decreases
with $N$).
For low values of $k$ (relative to fixed $N$)
we observe that the eigenvalue distribution
shows a peak at the center, presumably due to contributions from small
isolated clusters as mentioned in Ref. \cite{Far01}.
At higher values of $k$, we observe distributions similar to the ones
obtained for the constant $C$ case reported before.
However, a major difference was the relation between system size ($N$)
and the largest eigenvalue, $\lambda_{max}$, as well as the fraction of
real eigenvalues, $f_{real} ( \lambda )$ (Fig. 4). For random networks,
$\lambda_{max}$ attains a constant value for large values of $N$.
Further, as pointed out in Ref.~\cite{Som88}, the excess
density of real eigenvalues decreases with $N$ roughly as $N^{-1/2}$.
However, for regular networks, $\lambda_{max}$ grows with $N$ as ${\rm log}
( N^{\beta} )$ (in Fig. 4, $\beta \sim 0.07$) and the excess density of real
eigenvalues becomes constant for large $N$. This implies that as the
regular network increases in size, the tail of the eigenvalue distribution
gets longer (while the bulk remains fixed in size, similar to random networks).
Also, more and more eigenvalues migrate
from the complex plane to the real axis, keeping its density constant
even though the system size (and hence, the total number of eigenvalues)
is increasing. These results imply that for the case of fixed number of
links, regular networks are likely to be
more unstable than random networks as the system size is increased.

\vspace{-0.8cm}
\section{Conclusion}
\vspace{-0.8cm}
Based on the results reported above we conclude that
the introduction of small-world structure, i.e., a high degree
of clustering among the nodes of a network, does not increase the network
stability. On the contrary,
in certain conditions, such a structure might make the network more
unstable than its random counterpart. However, it was established
quantitatively (using finite-size scaling) that the nature of the 
stability-instability transition with increasing complexity appears
to change with the connection topology. In particular, networks with
higher degree of regularity destabilize more smoothly compared
to the abrupt transition to instability for random networks. This may
have implications for the occurrence of small-world structure in 
certain food webs. Although unable to make the network more stable,
such clustering structures may avoid the disastrous consequences of 
instability by making
the deterioration of the network more gradual, compared to a
random network.

I thank Ramesh Anishetty and Sanjay Jain for helpful discussions.

\end{document}